\documentclass{article}
\usepackage{amsmath}
\usepackage{amssymb}
\usepackage{amsbsy}
\usepackage{cite}

\usepackage{comment}
\usepackage{enumerate}
\usepackage{graphicx}

\newcommand{\be}{\begin{equation}}
\newcommand{\ee}{\end{equation}}

\def\bea{\begin{eqnarray}}
\def\eea{\end{eqnarray}}
\def\bean{\begin{eqnarray*}}
\def\eean{\end{eqnarray*}}

\newcommand{\barr}{\begin{array}}
\newcommand{\earr}{\end{array}}

\newcommand{\bed}{\begin{displaymath}}
\newcommand{\eed}{\end{displaymath}}
\newcommand{\bal}{\begin{array}{ll}}
\newcommand{\eal}{\end{array}}

\newcommand{\group}{$\mathcal{ PSL}_2(7)$ }
\newcommand{\cov}[3]{\big\{ {#1}\otimes{#2}\big\}^{}_{\bf #3}}
\newcommand{\scov}[2]{\big\{ {#1}\big\}^{}_{\bf #2}}
\newcommand{\pcov}[2]{\big[ {#1}\big]^{}_{\bf #2}}
\newcommand{\rep}[1]{\mathbf{#1}}

\def\bvec#1{\raise1.5ex\hbox{$\rightarrow$}\mkern-16.5mu #1}

\def\ket#1{\vert\,#1\rangle}
\def\m#1{\mathcal#1}

\newcommand{\bs}{\boldsymbol}
\newcommand\emb[2]{{\bf #1}_{\bf #2}}

\begin{document}

\title{\hfill ~\\[-30mm]
       \hfill\mbox{\small }\\[30mm]
       \textbf{ Neutrino masses, the $\mu$-term and \group}
       } 
\date{}
\author{\\ Gaoli Chen,\footnote{E-mail: {\tt gchen@ufl.edu}}~~M. Jay P\'erez,\footnote{E-mail: {\tt mjperez@phys.ufl.edu}}~~Pierre Ramond,\footnote{E-mail: {\tt ramond@phys.ufl.edu}}\\ \\
  \emph{\small{}Institute for Fundamental Theory, Department of Physics,}\\
  \emph{\small University of Florida, Gainesville, FL 32611, USA}}

\maketitle

\begin{abstract}
\noindent   
Using an $SO(10)$-inspired form for the Dirac neutrino mass, we map the neutrino data to right-handed neutrino Majorana mass-matrix, $\m M$, and investigate a special form with \emph{seesaw} tribimaximal mixing; it predicts a normal hierarchy, and the values of the light neutrino masses. It may be generated by mapping the top quark hierarchy onto the vacuum values of familon fields transforming under the family group $\mathcal{ PSL}_2(7)$. We next investigate the hypothesis that these familons play a dual role, generating a hierarchy in the supersymmetric $\bs \mu$-mass matrix of Higgses carrying family quantum numbers. A special $\mathcal{ PSL}_2(7)$ invariant coupling produces a $\mu$-matrix with a hierarchy of thirteen orders of magnitude. Only one Higgs field (per hypercharge sector) is light enough (with a $\mu$-mass $\sim 10-100$ GeV) to be destabilized by  SUSY soft breaking at the TeV scale, and upon spontaneous symmetry breaking, gives \emph{tree-level} masses for the heaviest family. 
\end{abstract} 

\thispagestyle{empty}
\vfill
\newpage
\setcounter{page}{1}
\section{Introduction}
The Standard Model, beautiful as it is, has many puzzling features:
\vskip .2cm
\noindent$\bullet$ The extreme hierarchy in the top quark sector \cite{ross&serna}.
\vskip .2cm
\noindent$\bullet$ The $\nu$-Standard Model contains the seesaw mechanism's  $\Delta I_{\rm w}=0$ Majorana matrix $\m M$ with unknown principles to determine its form \cite{seesaw}. 
\vskip .2cm
\noindent$\bullet$  In the MSSM, one finds a supersymmetric-invariant mass ($\mu$-term) of electroweak scale, indicating a conceptual puzzle \cite{martin}.
\vskip .2cm

In this paper, we suggest that these apparently disconnected facts can be addressed together through the orchestration of the discrete flavor symmetry $\mathcal {PSL}_2(7)$, a maximal discrete subgroup of $SU(3)$.  

The extreme hierarchy of the top quark sector presents us with a further mystery in Grand-Unified \cite{GUT,GUTrvw} models such as $SO(10)$ \cite{SO10rvw}. Minimal models of $SO(10)$ predict the relationship $Y^{(0)}_{} \sim Y^{(2/3)}$ \cite{so10}, implying the hierarchy of the charge-(2/3) quarks should appear in the Dirac neutrino mass spectrum; yet, according to the data \cite{dayabay,reno,dchooz,fit,Planck},  such a hierarchy is not reflected in the spectrum of the light neutrinos.

This tension can be relieved through the seesaw mechanism, in which the neutrino masses and mixing stem from a mixture of the Dirac neutrino mass matrix $\sim Y^{(0)}$ and the Majorana mass matrix of the right-handed neutrinos, $\m M$. In this case, $\m M$ must contain a squared-correlated hierarchy of its own. 

Beginning with the $SO(10)$-inspired assumption $Y^{(0)} = m_t \rm{diag}(\lambda^8,\lambda^4,1)$ and the neutrino oscillation data, we invert the seesaw mechanism and map the hierarchy of the top quark sector to $\m M$. Following some phenomenology, we find a special and elegant form for $\m M$: it predicts tribimaximal seesaw mixing angles, a normal hierarchy, and contains a Gatto-like relation, allowing us to determine the values of the light neutrino masses.

It is a finely-tuned matrix with precise relationships amongst its elements. However, searching for a natural way to produce $\m M$, we show it to be most simply obtained by a particular linear combination of two operators invariant under $\m T_7$, the smallest non-abelian finite subgroup of $SU(3)$. This combination is singled out when $\m T_7$ is enlarged to the group $\mathcal {PSL}_2(7)$. 

Its construction requires the introduction of two familon fields which carry the hierarchy of the top quark sector. In this paper, we {\em assume} their vacuum values as input, as required to reproduce $\m M$, although eventually they should be derived from a potential.

Having mapped the hierarchy of the top quark sector to the familon vacuum, we next investigate the hypothesis that these familons play a further role. 

We show they generate an acceptable hierarchy in the supersymmetric $\bs \mu$-mass matrix of a family-dependent Higgses. Indeed, we find a $\mathcal {PSL}_2(7)$-invariant coupling which produces a $\bs \mu$-matrix with a hierarchy of thirteen orders of magnitude. It is a solution of the $\mu$-term hierarchy, stemming from the $\mathcal {PSL}_2(7)$ Clebsch-Gordan coefficients and the hierarchy carried by the familon vacuum. It suggests a correlation between the overall scales of the mass matrices $\bs \mu$ and $\m M$. 

The eigenvectors of the $\bs\mu$-matrix show that the Higgs with the smallest $\mu$ mass couples to the $33$ component of the Yukawa matrices. If soft SUSY breaking terms are around the TeV scale, only the light Higgs in  $\m H_u$ and  $\m H_d$ will be affected and develop vacuum values, giving the correct pattern,
\bea
\left(\begin{array}{ccc}0 & & \cr & 0 & \cr & & 1\end{array}\right)
\nonumber
\eea
of tree-level masses for the quarks and leptons. 

Our tree-level mechanism does not account for the masses and mixings of the lightest two families, which will need to be  generated in an extension of this mechanism.

This is a proof of principle that ${\mathcal {PSL}_2(7)}$, a group with no doublet representations, can single out the third family. 

It is noteworthy that this pattern of  $\mu$-masses and mixings is determined by the {\em same} familon vacuum that is linked to the top quark hierarchy. 


\section{Preliminaries}

$SO(10)$ models in which the primary contribution to the quark and lepton masses arise through the coupling

$$ \bf{16} \cdot \bf{16} \cdot \bf{10}^{}_H $$
imply the relationship $Y^{(2/3)} \sim Y^{(0)}$: the mass hierarchy in the top quark sector, $(\lambda^8,\lambda^4, 1)$,  where $\lambda$ is the Cabibbo angle, should appear in the \emph{Dirac} neutrino mass spectrum.  The data shows no such hierarchy amongst the light neutrinos. 

The seesaw mechanism, natural in $SO(10)$, can relieve this tension through a right-handed neutrino Majorana mass matrix ${\m M}$, provided it contains a squared correlated hierarchy.  This is apparent by considering the seesaw relation between the light neutrino masses and $\m M$,

\be \label{eq:seesaw}
M^{}_\nu =  Y_{}^{(0)} \m M_{}^{-1} Y_{}^{(0) T},
\ee   
which with $Y^{(0)} \sim (\lambda^8,\lambda^4, 1)$ imply an eigenvalue pattern $(\lambda^{16}, \lambda^8, 1)$ for $\m M$.

We parametrize the hierarchy in $\m M$ by working in a basis where $Y^{(0)}$ is diagonal. Decompose $M_\nu$ as,

\begin{equation} \nonumber
M^{}_\nu = \m U^{}_{\rm \,seesaw}~ \m D^{}_\nu~ \m U^T_{\rm \,seesaw},
\end{equation}
where  $\m D_\nu={\rm diag}(m_1,m_2,m_3)$ is the diagonal neutrino mass matrix, and 
 $\m U_{\rm \,seesaw}$ is the Seesaw neutrino mixing matrix. We choose a standard parametrization for $\m U_{\rm seesaw}$, 
 
\be
{\m U}^{}_{\rm \,seesaw}~ = ~\begin{pmatrix}1&0&0\cr 0&c^{}_{23}&-s^{}_{23}\cr 0&s^{}_{23}&c^{}_{23}\end{pmatrix}
\begin{pmatrix}c^{}_{13}&0&-e^{-i\delta}_{}s^{}_{13}\cr 0&1&0\cr e^{i\delta}_{}s^{}_{13}&0&c^{}_{13}\end{pmatrix}
\begin{pmatrix}c^{}_{12}&-s^{}_{12}&0\cr s^{}_{12}&c^{}_{12}&0\cr 0&0&1\end{pmatrix},
\ee
written in terms of one Dirac CP-violating phase angle $\delta$ and rotation angles $\eta_{ij}$ ($s_{ij}=\sin\eta_{ij}$,   $c_{ij}=\cos\eta_{ij}$), to distinguish them from their measured counterparts $\theta_{ij}$ in the observable MNSP matrix, 
 
\be \nonumber
\m U^{}_{\rm MNSP}~=~\m U^\dagger_{-1}\,\m U^{}_{\rm seesaw},\ee 
with $\m U_{-1}$ determined by the charged lepton Yukawa matrix $Y^{(-1)}$. 

Eq.~\ref{eq:seesaw} may then be inverted and $\m M$ expressed in terms of the light-neutrino parameters,
 
\be \label{eq:hmaj}
{\m M}~\sim~\begin{pmatrix}a^{}_{11}\lambda^{16}_{}&a^{}_{12}\lambda^{12}_{}&a^{}_{13}\lambda^8_{}\cr a^{}_{12}\lambda^{12}_{}&a^{}_{22}\lambda^8_{}&a^{}_{23}\lambda^4_{}\cr a^{}_{13}\lambda^8_{}&a^{}_{23}\lambda^4_{}& a_{33} \end{pmatrix},\ee 
where the $a_{ij}$ are order one coefficients which depend on the light neutrino masses and seesaw mixing angles. 

Alternatively, the pre-factors $a_{ij}$ determine the low-energy neutrino parameters. This is particularly useful if the matrix of the fundamental theory is $\m M$, with $M_\nu$ resulting only after seesaw diagonalization; a point of view which we take here. 

A simple constraint among the pre-factors, ($2-3$ or $\mu-\tau$ symmetry \cite{mutau}), determines two of the three mixing angles appearing in $\m U_{\rm seesaw}$,

\begin{eqnarray} \label{eq:mutau}
a^{}_{12}=a^{}_{13},\qquad a^{}_{22}=a^{}_{33},~~~~\longrightarrow~~~~\eta^{}_{23}=45^\circ,\qquad \eta^{}_{13}~=~0^\circ,
\end{eqnarray}
with $\delta$, $\eta_{12}$  and the three masses undetermined.  The further relationship 

$$a^{}_{23}~=~ a^{}_{11}- a^{}_{12}- a^{}_{22} ~~~~\longrightarrow~~~~\tan^2\eta^{}_{12} = \frac{1}{2},$$
fixes the final mixing angle, providing tribimaximal \cite{TBM} \emph{seesaw} mixing,

\begin{eqnarray} \label{eq:TBM}
\m U_{\rm seesaw} = \m U_{\rm TBM} &= & \begin{pmatrix}
\sqrt{\frac{2}{3}} & -\frac{1}{\sqrt{3}} & 0 \cr
\frac{1}{\sqrt{6}} & \frac{1}{\sqrt{3}}  & -\frac{1}{\sqrt{2}} \cr
\frac{1}{\sqrt{6}} & \frac{1}{\sqrt{3}}  & \frac{1}{\sqrt{2}}
\end{pmatrix}.
\end{eqnarray}

As we now know $\theta_{13} \neq 0$, this seesaw mixing must be supplemented by the mixing matrix appearing in the diagonalization of the charged lepton Yukawa, $Y^{(-1)}$. In Grand-Unified theories such as $SU(5)$ where the down-type quarks and charged leptons Yukawas are related, this may be understood as a ``Cabbibo-Haze" \cite{haze} constraint: 

\begin{equation}
\m U^{}_{-1} ~=~ \begin{pmatrix} 1 & \lambda & 0 \cr -\lambda & 1 & 0 \cr 0 & 0 & 1\end{pmatrix} ~+~ \m O(\lambda^2) ~~\longrightarrow~~ \theta_{13} ~=~ \frac{\lambda}{\sqrt{2}} \approx 9^\circ.
\end{equation}

The further constraint \cite{Majorana_Paper}, 

\begin{equation}
a^{}_{23} ~=~ - a^{}_{22},
\end{equation}
produces a special and finely-tuned Majorana matrix,

\bea  \label{eq:smm}
{\m M } = M^{}_0 \begin{pmatrix}
r \lambda^{16} & r \lambda^{12} & r \lambda^{8} \cr r \lambda^{12} & \lambda^{8} & -\lambda^{4} \cr r \lambda^{8} & -\lambda^{4} & 1  
\end{pmatrix},\qquad r=\frac{m_{\nu_3}}{m_{\nu_1}}.
\eea
In a diagonal basis,  where $Y^{(0)}$ and $Y^{(2/3)}$ are along the $(\lambda^8, \lambda^4, 1)$ direction, it has a number of noteworthy attributes:

\begin{itemize}
\item The two lightest right-handed neutrino masses are almost degenerate
because of the vanishing of the sub-determinant in the $23$ block.

\item It predicts tribimaximal seesaw mixing.

\item The  parameter $r$ is determined from the data to be $\sim 10$, so that $\m M$ predicts a normal hierarchy among the light neutrino masses. 

\item It contains a relationship between the eigenvalues and mixing angles of $M_\nu$,

\be\label{Gatto}  \tan^2\eta_{12} ~=~ \left|\frac{m_{\nu_1}}{m_{\nu_2}}\right| ~=~\frac{1}{2},\ee 
reminiscent of the Gatto relation \cite{Go_Gatto!} in the quark sector, and where $\eta_{12}$ is the TBM solar neutrino mixing angle.

\end{itemize}
In addition, the extra relation of Eq.~\ref{Gatto}, $|m_{\nu_2}| = 2 |m_{\nu_1}|$, added to the neutrino oscillation data (for a review, see \cite{whitepaper}), yields the light neutrino masses,

\be
\label{eq:TBMprediction}
m_{\nu_1}^{} = 0.005 ~\textrm{eV} , \quad m_{\nu_2}^{} = 0.01 ~\textrm{eV}, \quad m_{\nu_3}^{} = 0.05 ~\textrm{eV},
\ee
as well as the mass of the heaviest right-handed neutrino,\footnote{This assumes an exact equality between $Y^{(0)}$ and $Y^{(2/3)}$. If the overall scale of $Y^{(0)}$ and $Y^{(2/3)}$ are different, this scale will be modified.} 

\be\label{scale}
M^{}_0~=~3\times 10^{14}_{} \rm{GeV},\ee
with the two ``light" right-handed neutrinos at $2r\lambda^{12}M_0\approx 10^8~GeV$.

Note that this matrix is capable of reproducing the light neutrino data only in the basis where the top quark lies in the $33$ position of the charge-$(2/3)$ quark Yukawa matrix. 
    
In an earlier work \cite{Majorana_Paper} we investigated whether this special form for $\m M$ could be naturally obtained from the family symmetry $\m T_7$. From a model building perspective, its most interesting feature is the vanishing of its sub-determinant in the $23$ block and the highly non-trivial relations among its elements. 
 

\section{The Special Majorana Matrix and $\m T_7$}

Our Ansatz for the special form of $\m M$ depends crucially on the hierarchy of the top quark sector, which is indicative of a tree-level pattern 

$$ 
Y \sim \begin{pmatrix} 0 & & \cr & 0 & \cr & & 1\end{pmatrix}
$$
for all Yukawa matrices. Such a pattern can be naturally accommodated in models with an $SU(3)$ (not $SO(3)$) family symmetry \cite{SU3}. Since the two large neutrino mixing angles suggest a crystallographic symmetry, we look amongst the finite subgroups of $SU(3)$. 

The smallest such subgroup of $SU(3)$, of order 21, is $\m T_7$. It contains five irreducible representations: a complex $\bf 3$ and its conjugate and three singlets $\bf 1$, $\bf 1^\prime$ and $\bf \bar{1}^\prime$. The relevant group theory for $\m T_7$ may be found in the Appendix A. 

One virtue of $\m T_7$ is that the covariants formed from products of two triplets, 

$$ {\bf 3}\otimes{\bf 3}= ({\bf 3}\oplus\bar{\bf 3})_s\oplus \bar{\bf 3}_a,$$
distinguish between diagonal and off-diagonal type couplings. Introducing three  right handed neutrinos, $N$, transforming as one $\m T_7$ triplet, the Clebsch-Gordan coefficients (see Appendix A) imply the covariant 

$$ 
\scov{ N  N}{3} 
$$
is a diagonal matrix, while 

$$ 
\scov{ N N }{\bar 3}
$$
is purely off-diagonal and symmetric since $N$ is Majorana. We introduce the \emph{curly-bracketed} $\left\{ \ldots \right\}$ notation to denote products of $\m T_7$ covariants.  As both diagonal and off-diagonal couplings are needed to produce $\m M$, at least two new \emph{familon} fields must be introduced which couple to the combinations $\scov{N N}{3}$ and $\scov{N N}{\bar 3}$. 

The special form of $\m M$, particularly its vanishing sub-determinant, requires precise relationships between its diagonal and off-diagonal elements. At the level of invariant couplings, this implies only \emph{particular} linear combinations of $\m T_7$ invariants are capable of producing the special Majorana matrix. When both familons are anti-triplets, we find $r$-independent linear combinations which can reproduce the special Majorana matrix.\footnote{When either or both familons are triplets, we find only $r$-dependent linear combinations, see Appendix B.}

Adding two familon anti-triplet fields $\bar \varphi$, $\bar \varphi^\prime \sim \bar{3}$, there are three covariant combinations which may couple to $\scov{NN}{}$,

$$ \scov{\bar \varphi \bar \varphi^\prime}{\bar 3}, \qquad \scov{\bar \varphi \bar \varphi^\prime}{3_s}, \qquad \scov{\bar \varphi \bar \varphi^\prime}{3_a}.$$
General $r$-independent solutions involves linear combinations of these three possible covariants. However, there exists a \emph{unique} linear combination involving only two invariants,  

\be \label{eq:lc} \scov{{ N} \,{ N }}{3} \scov{ \bar\varphi\bar\varphi' }{\bar 3}-
\scov{  N\, N }{\bar 3_s} \scov{ \bar\varphi\bar\varphi' }{3_s}.\ee

It is the simplest combination of dimension-five operators capable of producing $\m M$. 

It yields a  Majorana matrix of the form,

\bea
\label{eq:dim5}
\begin{pmatrix}
2 \bar{\varphi}_1\bar{\varphi}_1^\prime & -(\bar{\varphi}_1\bar{\varphi}_2^\prime + \bar{\varphi}_2\bar{\varphi}_1^\prime) & -(\bar{\varphi}_1\bar{\varphi}_3^\prime + \bar{\varphi}_3\bar{\varphi}_1^\prime) \cr
  -(\bar{\varphi}_1\bar{\varphi}_2^\prime + \bar{\varphi}_2\bar{\varphi}_1^\prime) & 2 \bar{\varphi}_2\bar{\varphi}_2^\prime & -(\bar{\varphi}_2\bar{\varphi}_3^\prime + \bar{\varphi}_3\bar{\varphi}_2^\prime) \cr
-(\bar{\varphi}_1\bar{\varphi}_3^\prime + \bar{\varphi}_3\bar{\varphi}_1^\prime) &  -(\bar{\varphi}_2\bar{\varphi}_3^\prime + \bar{\varphi}_3\bar{\varphi}_2^\prime)& 2\bar{\varphi}_3\bar{\varphi}_3^\prime
\end{pmatrix}.
\eea

In order for this coupling to yield the right-handed neutrino Majorana matrix, 

$$
{\m M}~=~M^{}_0
 \begin{pmatrix}
r \lambda^{16} & r \lambda^{12} & r \lambda^{8} \cr r \lambda^{12} & \lambda^{8} & -\lambda^{4} \cr r \lambda^{8} & -\lambda^{4} & 1  
\end{pmatrix}, 
$$
the vacuum values of the two antitriplet familons must be almost aligned, with \emph{linear} relations between the vacuum values of $\bar \varphi$ and $\bar \varphi^\prime$, 

\be \label{eq:famvac1}
\langle\bar{\varphi}\rangle = \langle \bar\varphi_3 \rangle \begin{pmatrix}
\bar{\alpha} \lambda^8 \cr  \lambda^4 \cr 1
\end{pmatrix}, \qquad \langle\bar{\varphi}^\prime\rangle= \langle \bar\varphi_3' \rangle  \begin{pmatrix}
\bar{\alpha}^\prime \lambda^8 \cr \lambda^4 \cr 1
\end{pmatrix},
\ee
where $ \bar{\alpha}\bar{\alpha}^\prime = - (\bar{\alpha}+\bar{\alpha}^\prime)/2 = r$. 

These vacuum values act as inputs to the model, and should eventually be derived from a potential.\footnote{One can build a familon potential which yields two aligned vacuum values, disturbed by a small perturbation which misaligns them ($\bar\alpha\neq \bar\alpha'$). In our bottom-up approach this misalignment constrains yet to be explored hidden sector physics.} Suffice it to say that a vacuum of the form, 

$$
\begin{pmatrix}
 0 \cr  0 \cr 1
\end{pmatrix}  
$$
occurs naturally in a $\m T_7$-invariant potential. Their near alignment, to order $\lambda^4$, is linked to the vanishing of the sub-determinant of $\mathcal{M}$.

While the necessary relative minus sign between independent couplings implies fine-tuning at the level of $\m T_7$, it can be natural if it arises from a particular symmetry. Such a symmetry is provided by the group $\mathcal{PSL}_2(7)$, of which $\m T_7$ is a subgroup. 

\section{From $\m T_7$ to \group}

The discrete group \group of order $168$ has six irreducible representations \cite{PSL,Atlas}: the complex $\bf 3$ and its conjugate, $\bf\bar 3$, as well as four reals, $\bf 1$, $\bf 6$, $\bf 7$, and $\bf 8$.\footnote{Much of the relevant group-theory is to be found in Appendix A.} The embedding of its subgroup \group $\supset \m T_7$, 

\begin{eqnarray} \nonumber
\bf 3 &=& \bf 3 \\ \nonumber
\bf \bar 3 &=& \bf \bar 3 \\ \nonumber
\bf 6 &=& \bf 3 \oplus \bf \bar 3 \\ \nonumber
\bf 7 &=& \bf 1 \oplus \bf 3 \oplus \bf  \bar 3 \\ \nonumber
\bf 8 &=& \bf 1^\prime \oplus \bf \bar 1^\prime \oplus 3 \oplus \bar 3,
\end{eqnarray}
shows that the triplets and anti-triplets of \group and $\m T_7$ may be identified. We therefore take our right-handed neutrinos to be triplets under \group.

The product of two \group triplets produces a sextet and an anti-triplet,

\be\label{Kron1}{\bf 3}\otimes{\bf 3}~=~{\bf 6_s}\oplus \bar{\bf{3}}_{\bs a}.\ee
The Majorana nature of the right-handed neutrinos then implies that the two $\m T_7$ bilinears $\scov{N N}{}$ make up the  \group sextet,

\be \pcov{N \ N}{6} ~=~ \scov{N \ N}{3} \oplus \scov{N \ N}{\bar 3_s},  \ee
where the \emph{square-brackets} $\left[ \ldots \right]$ denote \group covariants. Similarly, the symmetric product of the two anti-triplet familons transforms as a \group sextet,

\be \pcov{\bar \varphi ~ \bar \varphi^\prime }{6} ~=~ \scov{\bar \varphi ~ \bar \varphi^\prime }{\bar 3} \oplus \scov{\bar \varphi  ~ \bar \varphi^\prime }{ 3_s}.  \ee
As these are precisely covariants appearing in the special linear combination, it is tempting to search for Eq.~\ref{eq:lc} in the invariant product of two sextets,

\be \pcov{N \ N}{6} \cdot \pcov{ \bar \varphi ~ \bar \varphi^\prime}{6}. \ee 
This product shows two possibilities,

\be\label{Kron2}
{\bf 6}\otimes {\bf 6}~=~({\bf 1}\oplus {\bf 6}_1\oplus {\bf 6}_2\oplus {\bf 8})_s^{}\oplus ({\bf 7}\oplus {\bf 8})^{}_a.
\ee
One is the true \group singlet contained in the symmetric product, the other the $\m T_7$ singlet component of the septet lying in the antisymmetric product.   

The crucial minus sign points to the \group septet, as

\be \scov{\pcov{\pcov{N \ N}{6} \cdot \pcov{ \bar \varphi ~ \bar \varphi^\prime}{6}}{7}}{1} ~=~ \scov{{ N} \,{ N }}{3} \scov{ \bar\varphi\bar\varphi' }{\bar 3}-
\scov{  N\, N }{\bar 3_s} \scov{ \bar\varphi\bar\varphi' }{3_s}, \ee 
precisely the sought after linear combination. What was fine-tuned at the level of $\m T_7$ is naturally realized by \emph{one} $\mathcal{PSL}_2(7)$-invariant.

Having found a way to naturally produce $\m M$ through $\mathcal{PSL}_2(7)$ familons which carry the hierarchy of the top-quark sector in their vacuum values, an intriguing question is whether these same familons can be used for another purpose. We next investigate the hypothesis that they are responsible for a hierarchy amongst a $\mathcal{PSL}_2(7)$ family of Higgses.

\subsection*{$\mathcal{PSL}_2(7)$ Higgses}

We assign all matter fields to $\mathcal {PSL}_2(7)$-triplets. The invariant tree-level Yukawa couplings are determined from the product,

$${\bf 3}\otimes{\bf 3}~=~{\bf 6}\oplus \bar{\bf 3},$$
which resembles $SU(3)$, except that the $\bf 6$ is real \cite{SU3}. We choose the Higgs fields to transform as a $\mathcal{PSL}_2(7)$ sextet.\footnote{Although the Kronecker product allows them, we do not include tree-level $\mathcal{PSL}_2(7)$ anti-triplets which only contribute to antisymmetric Yukawa matrices.} The $\m T_7$ decomposition gives,

\be
{\bf 6}={\bf 3}\oplus{\bf\bar 3},\qquad    {{\m H}^{}_{u,d}}~=~ { H}^{}_{u,d} \oplus {\overline H}^{}_{u,d},
\ee
where $\m H_{u,d}$ are $\mathcal {PSL}_2(7)$ sextets and $H_{u,d}$ and $\overline H_{u,d}$ are $\m T_7$ triplets and anti-triplets, respectively. The sextet coupling contains both diagonal and symmetric off-diagonal Yukawa couplings,

\be 
\pcov{Q \ \bar u}{6} \m H^{}_{u} ~=~ \scov{Q \bar u}{3} \overline H_{u \bf} + \scov{Q \bar u}{\bar 3} H_{u}.
\ee
The first term is diagonal, capable of producing the tree-level top quark mass provided the first component of the anti-triplet

\be \langle \overline H_{u 1} \rangle_0 \neq 0, \ee
does not vanish in vacuum. As our goal is to produce the correct tree-level pattern of Yukawa matrices, we will take our Higgs fields to be \group sextets. 

Although we have found a coupling to generate the top quark mass, it comes at the cost of a large number of Higgs fields. Only one of these (per hypercharge) should be light and play the role of the MSSM Higgs. We now show that with a special choice of \group couplings, the familons of the Majorana sector may be used to produce an acceptable splitting between a light Higgs and its family partners.

\section{$\mathcal {PSL}_2(7)$-invariant Theory}

We introduce two sextet Higgs fields ${\m H}^{}_u$ and ${\m H}^{}_d\sim{\bf 6}$, which couple to the matter fields as 

\be
{\m W}^{}_Y~=~ \pcov{{ Q}{\bar u}}{6} {\m H}^{}_{u}~+~\pcov{{ L}{ N}}{6}{ \m H}^{}_{u}~+~ \pcov{{Q}{\bar d}}{6}{\m H}^{}_{d}~+~\pcov{{ L}{\bar e}}{6}{ \m H}^{}_{d},
\ee
Dimensionless couplings, assumed to be of order one, are not displayed. To this Yukawa superpotential, we add the $\mathcal {PSL}_2(7)$-invariant ``familon superpotential", 

\be \label{eq:fampot}
{\m W}^{}_{\rm Fam}~=~\Big(\pcov{{N}\,{N}}{6}+f\pcov{{\m H^{}_{u}}{\m H^{}_{d}}}{6_1} \Big)\Phi+\pcov{\bar\varphi\bar\varphi'}{6}\bar\Phi+\Phi{\bar \Phi}S^{}_{\bf 7},\ee
where $f$ is an arbitrary parameter. The fields $\bar\varphi,~\bar\varphi'\sim\bar 3$ are the anti-triplets of Eq.~\ref{eq:famvac1}, while $\Phi,~\bar\Phi\sim{\bf 6}$, and $S^{}_{\bf 7}\sim{\bf 7}$  are familon fields with only \group quantum numbers. They  induce non-renormalizable interactions which  generate $\m M$ and the $\mu$-term for the Higgs fields. 

The $\bf 6_1$ covariant appearing in the superpotential is one of the two sextets in the \group Kronecker product,

\be\label{Kron3}
{\bf 6}\otimes {\bf 6}~=~({\bf 1}\oplus {\bf 6}_1\oplus {\bf 6}_2\oplus {\bf 8})_s^{}\oplus ({\bf 7}\oplus {\bf 8})^{}_a.
\ee
They are not distinguished by $\mathcal {PSL}_2(7)$, however $\bf 6_1$ is the one which appears explicitly in the $SU(3)$ Kronecker product \cite{deSwart},
 
 \be
 \bar{\bf 6}\otimes \bar{\bf 6}~=~(\overline{\bf 15}'\oplus {\bf 6})_s^{}+\overline{\bf 15}_a^{},
 \ee
while ${\bf 6}_2$ is part of the symmetric fourth rank tensor ($\overline{\bf 15}'$). Our choice of coupling reflects the $SU(3)$ origin of the theory. 

$\m W_Y$ implies three global symmetries: total lepton number $\m L$, baryon number $\m B$, and the $\m {PQ}$  chiral symmetry (normalized to $ \m {PQ}_{\m H_u}= \m {PQ}_{\m H_d}=1$) which forbids a tree-level $\mu$ term. These are further restricted by the familon structure. In addition to  $R$-symmetry, the full superpotential  $\m W_Y+\m W_{\rm Fam}$ supports three global symmetries shown in Table 1. Note that the lepton number and Peccei-Quinn symmetries of the Yukawa sector are linked, leaving the combination $\m L'=\m L-\frac{2}{3}{\m PQ},$ unbroken. 

To avoid Nambu-Goldstone bosons, further $\mathcal {PSL}_2(7)$-invariant terms built out of familons are required to provide  explicit symmetry breakings.

\begin{table}
\centering
\begin{tabular*}
{\textwidth}
{c c c c c c c c}
\hline
\hline
\noalign{\smallskip}
~~~~$U(1)$~~~~&$ N$&~~~~~ $\m H_{u,d}$&~~~~~$\Phi$& ~~~~~$\bar\Phi$  &~~~~~  $\bar\varphi$ &~~~~~ $\bar\varphi'$ &~~~~~ $S_{\bf 7}$\\
\noalign{\smallskip}
\hline
\noalign{\smallskip}\noalign{\smallskip}
$\m L'$ & $-\frac{2}{3}$&$-\frac{2}{3}$ & ~~~~~$\frac{4}{3}$ & $~~~~-\frac{4}{3}$&~~~~~~$\frac{2}{3}$ &~~~~~$\frac{2}{3}$&~~~~~$0$\\
\noalign{\smallskip}
\noalign{\smallskip}
$\m X$ & $-\frac{2}{3}$&$-\frac{2}{3}$ & ~~~~~$\frac{4}{3}$ & ~~~~~~$\frac{4}{3}$& $~~~~-\frac{2}{3}$& ~~~~$-\frac{2}{3}$&~~~~~$-\frac{8}{3}$\\
\noalign{\smallskip} \noalign{\smallskip}
$\m X'$ &~~$0$& ~~~$0$ & $~~~~~0$ &~~~~~~ $0$&~~~~~~ $1$&~~~~ $-1$&~~~~~$0$\\
\hline
\end{tabular*}
\caption{Global Symmetries of the Yukawa and Family Superpotentials.}
\end{table} 
  

\group is spontaneously broken to its $\m T_7$ subgroup by assuming that  $S_{\bf 7}$ gets a vev, since ${\bf 7}={\bf 1}+{\bf 3}+\bar{\bf 3}$ contains one $\m T_7$ singlet. It generates  a vector-like mass term for  the $\Phi,~\bar\Phi$ fields of the order of \group breaking. 

Integrating over these heavy fields produces the sought after $\m T_7$-invariant effective interaction,

\be \scov{{ N} \,{ N }}{3} \scov{ \bar\varphi\bar\varphi' }{\bar 3}-
\scov{  N\, N }{\bar 3_s} \scov{ \bar\varphi\bar\varphi' }{3_s},\ee
The crucial minus sign stemming from the coupling to the septet. Its evaluation in the familon vacuum of Eq.~\ref{eq:famvac1} yields the special Majorana matrix, with the overall scale

\be\label{scale1}
M^{}_0~=~\frac{\langle \bar\varphi_3 \rangle\langle \bar\varphi_3' \rangle}{M_\Phi}
,\ee
for the mass of the heaviest right-handed neutrino, and $M_\Phi$ is the mass of $\Phi$ and $\bar\Phi \sim \langle S_7 \rangle$. 

Using Eq.~\ref{scale1},  and the requirement that $M_\Phi$ be large,  we conclude that the familon  vacuum values must  themselves be large  and well above the TeV scale. Therefore we do not expect the familons to play a role in low energy phenomenology.
\vskip .3cm
\noindent {\large \bf The $\bs\mu$-term}
\vskip .3cm
\noindent The MSSM's $\mu$-term is generated through the coupling of the electroweak-invariant ${\m H^{}_{u}}{\m H^{}_{d}}$ to familons. Integration over the heavy $\Phi$ and $\bar \Phi$ fields generate the effective interaction,
 
\be\label{muterm-61}
\frac{f}{M_\Phi} \pcov{{\m H^{}_{u}}{\m H^{}_{d}}}{6_1}\pcov{\bar\varphi\bar\varphi'}{6},
 \ee
 which expressed in terms of $\m T_7$ covariants,  gives
 
 \bea
&&\frac{f}{M^{}_\Phi\sqrt{6}}\Big( \scov{ H^{}_{u}H^{}_{d} }{3}
-\sqrt{2} \scov{\overline H^{}_{u}\overline H^{}_{d} }{3_s}
 \Big) \cdot \scov{ \bar\varphi\bar\varphi' }{\bar 3}-\nonumber\\ ~~~ ~~~&&-
 \frac{f}{M^{}_\Phi\sqrt{6}} \Big( \scov{ H^{}_{u}H^{}_{d} }{{\bar 3}_s}
 - \scov{ H^{}_{u}\overline H^{}_{d} }{\bar 3}- \scov{ \overline H^{}_{u} H^{}_{d} }{\bar 3}\Big) \cdot \scov{ \bar\varphi\bar\varphi' }{3_s}.
 \eea 
Its evaluation in the familon vacuum of Eq.~\ref{eq:famvac1} yields the $\bs\mu$ matrix,
 
\be
{\m H}^T_u~{\large {\bs\mu}}~{\m H}^{}_d
\ee
where  
\be
{\large {\bs\mu}}~=~{M'}^{}_0\,\begin{pmatrix}
-r\lambda^{16} &r\lambda^{12} & r\lambda^{8}&0&\sqrt{2}\lambda^4&0 \cr
  r\lambda^{12} & -\lambda^{8} &-\lambda^{4}&0&0&-\sqrt{2}r\lambda^8 \cr
r\lambda^8 &  -\lambda^4& -1&-\sqrt{2}r\lambda^{12}&0&0\cr
0&0&-\sqrt{2}r\lambda^{12}&0&\lambda^8&r\lambda^{16}\cr
\sqrt{2}\lambda^4&0&0&\lambda^8&0&1\cr
0&-\sqrt{2}r\lambda^8&0&r\lambda^{16}&1&0\end{pmatrix},
\ee
and using Eq.~\ref{scale},

\be
{M'}^{}_0=\frac{f}{3\sqrt{2}}M^{}_0~=~\frac{f}{\sqrt{2}}\times 10^{14}_{}~GeV.
\ee
The $\mu$ values for the six Higgs fields are, up to the dimensionless parameter $f$,  determined by the neutrino masses and the hierarchy in the top quark sector.
 
A suitable rotation diagonalizes this matrix and provides the masses of the six Higgs fields in each sector (see Appendix C for details):

\begin{itemize}

\item Three Higgs fields per hypercharge, $E^{(1)}_{u,d}$, $E^{(2)}_{u,d}$,  $E^{(3)}_{u,d}$  with almost degenerate $|\mu|=|{M'}^{}_0|(1,1,1)$, respectively:  

\item Two Higgs fields per hypercharge $D^{(1)}_{u,d}$, $D^{(2)}_{u,d}$, with smaller $|\mu|=2r\lambda^{12}|{M'}^{}_0|\sim 10^8~GeV$

\item One Higgs field $h_{u,d}$ with $\mu=8r^2\lambda^{24}{M'}^{}_0= 27\,f~ GeV.$

\end{itemize}  
 
Through the orchestration of the flavor symmetry $\mathcal {PSL}_2(7)$, the Higgs spectrum parallels that of the right-handed neutrino masses, except for  $h_{u,d}$, which have $\mu$-values of the order of the electroweak scale. 

This provides an alternative answer for the anomalously low value of the $\mu$-term in the MSSM. We assume $f$ is of order one; experimental bounds on the $\mu$-term favors $f\ge 3$. 

The scale of supersymmetry breaking determines which of these Higgs fields can develop a vacuum value. With the traditional TeV-scale SUSY soft breaking terms, as implied by the convergence of the gauge coupling constants, only $h_{u,d}$  can have vacuum values and serve as the BEH scalars of the MSSM. The other five Higgs will not develop vacuum values because their $\mu$ mass is much larger than the soft breaking scale. 

Recall that in the basis set by our analysis of the seesaw masses, the top quark mass is generated by the vacuum value of ${\overline H}^{}_{u1}$ which couples to $Q^{}_3{\bar u}_3^{}$. On the other hand, it is the eigenvectors (see Appendix C)  of the $\bs\mu$-matrix  which determine which quarks get masses from ${\overline H}^{}_{u1}$'s vacuum value. 

The $\bs\mu$-matrix eigenvector belonging to $h_{u}$ is given by, 

\be
h_u~=~-\overline H_{u1}+\sqrt{2}\lambda^4\,H^{}_{u1}+ \cdots
\ee
Upon the electroweak breaking, it induces the tree-level masses,

\be
m^{}_t\begin{pmatrix}r(2r-1)\lambda^{16}&-3r\lambda^{12}&r\lambda^8\cr
-3r\lambda^{12}&\lambda^8&-\lambda^4\cr
r\lambda^8&-\lambda^4&1\end{pmatrix},
\ee
with the top quark tree-level mass in the correct position. Its eigenvalues are,

\be
m^{}_t,\qquad 2r\lambda^{12}m^{}_t,\qquad-2r\lambda^{12}m^{}_t,\ee
which shows degenerate tree-level $u$ and $c$ quark masses much below their actual values. This is tree-level result, and a more complete model should generate them either through higher order operators or by radiative correction. 

These results, a hierarchy of twelve orders of magnitude and a light Higgs along the right direction to produce the top quark mass, depend crucially on the Clebsch-Gordan coefficients of $\mathcal{PSL}_2(7)$. 

Had we chosen the $\bs 6_2$ coupling for $\m H_u \m H_d$,

\be \frac{f}{M_{\Phi}}\left[\mathcal{H}_{u}\mathcal{H}_{d}\right]_{6_{2}}\left[\bar{\varphi}\bar{\varphi}^{\prime}\right]_{6}, \ee
the $\bs \mu$-matrix would have been

\be 
{\large {\bs\mu}} ~=~ \frac{f}{3\sqrt{7}}M_{0}\left(\begin{array}{cccccc}
r\lambda^{16} & 2r\lambda^{12} & 2r\lambda^{8} & 0 & -\sqrt{2}\lambda^{4} & -\frac{3\lambda^{8}}{\sqrt{2}}\\
2r\lambda^{12} & \lambda^{8} & -2\lambda^{4} & -\frac{3}{\sqrt{2}} & 0 & \sqrt{2}r\lambda^{8}\\
2r\lambda^{8} & -2\lambda^{4} & 1 & \sqrt{2}r\lambda^{12} & -\frac{3r\lambda^{16}}{\sqrt{2}} & 0\\
0 & -\frac{3}{\sqrt{2}} & \sqrt{2}r\lambda^{12} & -3r\lambda^{8} & \frac{\lambda^{8}}{2} & \frac{r\lambda^{16}}{2}\\
-\sqrt{2}\lambda^{4} & 0 & -\frac{3r\lambda^{16}}{\sqrt{2}} & \frac{\lambda^{8}}{2} & -3r\lambda^{12} & \frac{1}{2}\\
-\frac{3\lambda^{8}}{\sqrt{2}} & \sqrt{2}r\lambda^{8} & 0 & \frac{r\lambda^{16}}{2} & \frac{1}{2} & 3\lambda^{4}
\end{array}\right).
\ee 
At leading order, its eigenvalues,

$$
\frac{f}{3\sqrt{7}}M_{0}\left\{ 1,-\frac{1}{2},\frac{1}{2},-\frac{3}{\sqrt{2}},\frac{3}{\sqrt{2}},12\lambda^{12}\right\},$$
display a milder hierarchy. With the $M_0$ of Eq.~\ref{scale}, the lightest Higgs, of $\mu$-mass $\frac{4f}{\sqrt{7}}M_{0}\lambda^{12}\approx f\times10^{7}\mathrm{GeV}$, is too heavy to be destabilized by Soft-Susy breaking effects and will not acquire a vacuum value. 

If we lower its mass either by changing $M_0$ or $f$, $h_{u}$ will acquire a vacuum value, but in the wrong direction. The $h_u$ eigenvector is given by,

\be
H_{u1}-3\lambda^{12}H_{u2}-2r\lambda^{8}H_{u3}+\frac{10}{3}\sqrt{2}r\lambda^{12}\overline{H}_{u1}-9\sqrt{2}\lambda^{8}\overline{H}_{u2}+2\sqrt{2}\lambda^{4}\overline{H}_{u3},
\ee
mostly made up of $H_{u1}$ and would induce a tree-level mass matrix proportional to,

\be
\left(\begin{array}{ccc}
0 & 0 & 0\\
0 & 0 & 1\\
0 & 1 & 0
\end{array}\right) + \m O(\lambda^4),
\ee
 which implies degenerate charm and top quark masses. This shows our mechanism could have failed in two different ways, either by not producing a light enough Higgs or by putting the light Higgs in wrong direction. As remarked earlier, this distinction between the $\bs 6_1$ and $\bs 6_2$ may hint at an underlying $SU(3)$ origin of the family symmetry.

\newpage
\section{Conclusions}
We have presented a mechanism based on the flavor symmetry $\mathcal{ PSL}_2(7)$, in which a particular Majorana matrix appears naturally\footnote{A review of flavor models is given in \cite{flavor}.}. It entails the existence of familons whose vacuum values generate the hierarchy of the top quark sector. 

From this Majorana matrix, and using the neutrino oscillation data, the masses of the right-handed neutrinos, as well as those of the three light neutrinos and their (TBM) mixing are determined: the heaviest right-handed neutrino is at $10^{14}$ GeV, while the two lighter ones weigh  $10^8$ GeV.  

The Higgs fields $\m H_{u,d}$ transform as sextets under the family symmetry. With the coupling of Eq.~\ref{muterm-61}, we found that their $\mu$-masses span a hierarchy of thirteen orders of magnitude. Three Higgs have $\mu$-mass $f\times 10^{14}$ Gev, two at $f\times 10^8$ GeV, and one light Higgs at $\sim 30f$ GeV commensurate with the electroweak   
scale. An  $f$ of order one poses no inconsistencies. This provides a solution to the $\mu$-term problem.

Soft SUSY-breaking around the TeV scale will not affect the five heavy Higgs fields in each hypercharge sector and only the light Higgs will get a vacuum value. It is aligned as to give tree-level masses to the third family of quarks and charged leptons, but does not account for the masses of the two lightest families. Our mechanism must therefore be supplemented by adding radiative structure and possible higher-dimensional operators.

Our mechanism could have gone wrong in many ways. It relies on the special properties of the ${\bf 6}_1$ coupling which yields a grand-unified type hierarchy, produces an unsuppressed tree-level top quark mass, and points to the continuous $SU(3)$. On the other hand, the ${\bf 6}_2$ coupling produces a milder hierarchy, with two light Higgs per hypercharge. In this case, $f$ would have to be very small for these fields to acquire vacuum values upon Susy breaking. In addition, the eigenvectors of the $\bs\mu$-matrix for this coupling are aligned in the wrong direction and produce degenerate charm and top quark masses.

The hierarchies in the seesaw and $\mu$-term couplings are generated by the top quark sector hierarchy. Surprisingly, they are close to the Grand-Unified hierarchies, suggesting a deeper relation between GUT and familon physics.

Although we have structured our model along the lines of $SO(10)$, motivating the relationship between the charge-(2/3) and neutral Dirac Yukawa matrices as well as the spinor-like representation assignment of our matter fields under the family symmetry, the gauge structure of our model at this stage is that of the Standard Model. In a future publication, we will consider an $SO(10)$ GUT extension of our model. 


Our approach has been in the ``bottom-up" mode, and we have not looked for a theoretical framework in which \group is natural. It may point to a continuous $SU(3)$ family symmetry as many authors have suggested.  \group is also a group ubiquitous in mathematics: it is the maximal discrete subgroup of $M_{21}$ and the group of symmetries acting on the Klein quartic surface, to name a few. 
\section{Acknowledgements}
We thank Jue Zhang, James Gainer, and Steve Martin for their helpful discussions, and especially to Tom Kephart for his careful reading of this manuscript. One of us (PR) wishes to thank the Aspen Center for Physics (partially supported by NSF grant PHY-106629), where part of this work was performed. MP would like to thank the McKnight Doctoral Fellowship Program for their continued support. This research is partially supported by the Department of Energy Grant No. DE-FG02-97ER41029.
 

\newpage
\appendix
\numberwithin{equation}{section}

\section{$\m T_7$ and \group Group Theory}
   
\noindent {\large\bf Kronecker Products of $\m T_7$ representations}
\vskip .3cm
\noindent The Kronecker products of triplets and antitriplets are given as
$$ {\bf 3}\otimes{\bf 3}= ({\bf 3}\oplus\bar{\bf 3})_s\oplus \bar{\bf 3}_a,\qquad {\bf 3}\otimes \bar{\bf 3}={\bf 1}\oplus {\bf 1}'\oplus \bar{\bf 1'}\oplus {\bf 3}\oplus \bar{\bf 3},
$$ \\
where $s$ ($a$) refers to symmetric (antisymmetric) part of the product. We can build $\m T_7$ covariants from these Kronecker products. For example, the first Kronecker product gives three $\m T_7$ covariants: $\{{\bf 3}\otimes{\bf 3}\}_{3}$, $\{{\bf 3}\otimes{\bf 3}\}_{\bar{3}_{s}}$, and $\{{\bf 3}\otimes{\bf 3}\}_{\bar{3}_{a}}$, where the subscript denotes the representation of the covariant. This notation is used in the text and the remaining parts of the Appendix.


\vskip .5cm
\noindent {\large\bf  $\m T_7$ Clebsch-Gordan coefficients}
\vskip .5cm
\bean
({\bf 3}\otimes {\bf 3^\prime})_s^{}~&\longrightarrow&~~{\bf 3}:~~\begin{cases}\ket 3\ket{3'}\cr \ket 1\ket{1'}\cr\ket 2\ket{2'}\end{cases}\ ;\qquad ~~\longrightarrow~~{\bf\overline 3}_s:~~\begin{cases}\frac{1}{\sqrt{2}}\left(\ket 3\ket{2'}+~\ket2\ket{3'}\right)\cr\frac{1}{\sqrt{2}}\left(\ket 1\ket{3'}+~\ket 3\ket{1'}\right)\cr\frac{1}{\sqrt{2}}\left(\ket 2\ket{1'}+~\ket1\ket{2'}\right)\end{cases}\;\\
\noalign{\bigskip}({\bf 3}\otimes {\bf 3^\prime})_a^{}~&\longrightarrow&~~ {\bf\overline 3}_a:~~\begin{cases}\frac{1}{\sqrt{2}}\left(\ket 3\ket{2'}-~\ket2\ket{3'}\right)\cr\frac{1}{\sqrt{2}}\left(\ket 1\ket{3'}-~\ket 3\ket{1'}\right)\cr\frac{1}{\sqrt{2}}\left(\ket 2\ket{1'}-~\ket1\ket{2'}\right)\end{cases}\ .
\eean
\vskip .5cm
\bean
{\bf 3}\otimes {\bf\overline 3}~&\longrightarrow~~&{\bf 3}:~~\begin{cases}\ket 2\ket{\overline  1}\cr \ket 3\ket{\overline  2}\cr\ket 1\ket{\overline  3}\end{cases}\ ;\qquad\qquad ~\longrightarrow~~{\bf\overline 3}:~~\begin{cases}\ket 1\ket{\overline  2}\cr \ket 2\ket{\overline  3}\cr\ket 3\ket{\overline  1}\end{cases}\ ,\\
&&\\
\noalign{\bigskip}{\bf 3}\otimes {\bf\overline 3}~&\longrightarrow~~&{\bf 1{\phantom{'}}}:~~\frac{1}{\sqrt{3}}\,\left( \ket 1\ket{\overline  1}+ ~\ket
  2\ket{\overline  2}+ ~ \ket 3\ket{\overline  3}\right)\ , \nonumber \\ 
\noalign{\bigskip}{\bf 3}\otimes {\bf\overline 3}~&\longrightarrow~~&{\bf 1'}:~~\frac{1}{\sqrt{3}}\,\left( \ket 1\ket{\overline  1}+ ~\omega^2_{}\ket
 2\ket{\overline  2}+ ~\omega\:\ket 3\ket{\overline  3}\right)\ ,~~~~~~~~~~~~~
 \\
\noalign{\bigskip}{\bf 3}\otimes {\bf\overline 3}~&\longrightarrow~~&{\bf \overline 1'}:~~\frac{1}{\sqrt{3}}\,\left( \ket 1\ket{\overline  1}+ ~\omega\:\ket  2\ket{\overline  2}+~ \omega^2_{}\ket 3\ket{\overline  3}\right)\ ,~~\omega=\exp(2i\pi/3) \\ \\ {\bf 1^\prime}\otimes{\bf 3}~&\longrightarrow&~~ {\bf 3}:~~\begin{cases}s^\prime \ket 1 \cr s^\prime \omega \ket 2 \cr s^\prime \omega^2 \ket 3\end{cases} \\ 
\eean

\newpage
\vskip .5cm
\noindent {\large\bf Decompositions of  \group irreps under $ \m T_7$}
\vskip .3cm
\begin{eqnarray*}
{\bf 3} & = & {\bf 3},\quad \bar{\bf 3} ~=~ \bar{\bf 3},\quad
{\bf 6} ~=~ {\bf 3}\oplus \bar{\bf 3},\\ 
\noalign{\bigskip}
{\bf 7} &=& {\bf 1}\oplus{\bf 3}\oplus \bar{\bf 3},\quad
{\bf 8} ~=~ {\bf 1'}\oplus \bar{\bf 1}'\oplus {\bf 3}\oplus \bar{\bf 3}.
\end{eqnarray*}

\vskip .5cm
\noindent {\large\bf Kronecker Products of  \group representations}
\vskip .3cm
\begin{eqnarray*}
{\bf 3}\otimes{\bf 3} &=& \bar{\bf 3}_{a}\oplus{\bf 6}_{s}, \qquad~~~~ {\bf 3}\otimes\bar{\bf 3} = {\bf 1}\oplus{\bf 8} \\
\\
{\bf 3}\otimes{\bf 6} &=& \bar{\bf 3}\oplus {\bf 7}\oplus{\bf 8},\qquad~~ \bar{\bf 3}\otimes{\bf 6}={\bf 3}\oplus {\bf 7}\oplus{\bf 8} \\
{\bf 3}\otimes{\bf 7} &=& {\bf 6}\oplus {\bf 7}\oplus{\bf 8},\qquad~~ \bar{\bf 3}\otimes{\bf 7}={\bf 6}\oplus {\bf 7}\oplus{\bf 8} \\
{\bf 3}\otimes{\bf 8} &=& {\bf 3}\oplus {\bf 6}\oplus {\bf 7}\oplus{\bf 8},\quad \bar{\bf 3}\otimes{\bf 8}=\bar{\bf 3}\oplus{\bf 6}\oplus {\bf 7}\oplus{\bf 8} \\
\\
{\bf 6}\otimes{\bf 6} &=& \left({\bf 1}\oplus{\bf 6}\oplus{\bf 6}\oplus{\bf 8}\right)_{s}\oplus\left({\bf 7}\oplus{\bf 8}\right)_{a} \\
{\bf 6}\otimes{\bf 7} &=& {\bf 3} \oplus \bar{\bf 3} \oplus {\bf 6} \oplus {\bf 7} \oplus{\bf 7} \oplus{\bf 8} \oplus{\bf 8} \\
{\bf 6}\otimes{\bf 8} &=& {\bf 3} \oplus \bar{\bf 3} \oplus {\bf 6} \oplus {\bf 6} \oplus {\bf 7} \oplus{\bf 7} \oplus{\bf 8} \oplus{\bf 8} \\
\\
{\bf 7}\otimes{\bf 7} &=& \left({\bf 1}\oplus{\bf 6}\oplus{\bf 6}\oplus{\bf 7}\oplus{\bf 8}\right)_{s}\oplus\left({\bf 3} \oplus \bar{\bf 3}\oplus{\bf 7}\oplus{\bf 8}\right)_{a} \\
{\bf 7}\otimes{\bf 8} &=& {\bf 3} \oplus \bar{\bf 3} \oplus {\bf 6} \oplus {\bf 6} \oplus {\bf 7} \oplus{\bf 7} \oplus{\bf 8} \oplus{\bf 8}\oplus{\bf 8} \\
\\
{\bf 8}\otimes{\bf 8} &=& \left({\bf 1}\oplus{\bf 6}\oplus{\bf 6}\oplus{\bf 7}\oplus{\bf 8}\oplus{\bf 8}\right)_{s}\oplus\left({\bf 3} \oplus \bar{\bf 3}\oplus{\bf 7}\oplus{\bf 7}\oplus{\bf 8}\right)_{a}
\end{eqnarray*}

\vskip .5cm
\noindent {\large\bf \group Clebsch-Gordan Coefficients}
\vskip .2cm

\noindent The Clebsch-Gordan coefficients used in the text are expressed in terms of  $\m T_7$ covariants. A \group sextex ${\bf 6}$ is decomposed as 
\begin{eqnarray*}
{\bf 6} \equiv {\emb 3 6} \oplus {\emb {\bar 3} 6},
\end{eqnarray*}
where ${\emb 3 6}$ (${\emb {\bar 3} 6}$) denotes the $\m T_7$ triplet (antitriplet) inside the \group ${\bf 6}$. The same applies for another sextex ${\bf 6^{\prime}}$,
\begin{eqnarray*}
{\bf 6^{\prime}} \equiv {\emb 3 {6^\prime}} \oplus {\emb {\bar 3} {6^\prime}}.
\end{eqnarray*}
The Clebsch-Gordan coefficients are:

\begin{itemize}
\item $\mathbf{6}\otimes\mathbf{6}^{\prime}\to\mathbf{1}:$
\[
\mathbf{1}=\frac{1}{\sqrt{2}}\cov{\emb 3 6}{\emb {\bar 3} {6^{\prime}}}{1}+\frac{1}{\sqrt{2}}\cov{\emb {\bar 3} 6}{\emb 3 {6^{\prime}}}{1}
\]

\item $\left(\mathbf{6}\otimes\mathbf{6}^{\prime}\right)_{s}\to\mathbf{6}_{1}\oplus\mathbf{6}_{2}:$
\begin{eqnarray*}
\mathbf{6}_{1} & = & \frac{1}{\sqrt{3}}\cov{\emb 3 6}{\emb 3 {6^{\prime}}}{3}-\sqrt{\frac{2}{3}} \cov{\emb {\bar 3}  6}{\emb {\bar 3} {6^{\prime}}}{3_{s}}- \\
& & -\frac{1}{\sqrt{3}}\cov{\emb 3 6}{\emb 3 {6^{\prime}}}{\bar{3}_{s}} + 
\frac{1} {\sqrt{3}}\cov{\emb 3 6}{\emb {\bar 3} {6^{\prime}}}{\bar 3} +
\frac{1}{\sqrt{3}}\cov{\emb {\bar 3} 6}{\emb 3 {6^{\prime}}}{\bar 3} \\
\mathbf{6}_{2} & = & -\sqrt{\frac{2}{21}}\cov{\emb 3 6}{\emb 3 {6^{\prime}}}{3}-\frac{1}{\sqrt{21}}\cov{\emb  {\bar3} 6}{\emb {\bar3} {6^{\prime}}}{3_{s}} + \sqrt{\frac{3}{7}}\cov{\emb 3 6}{\emb {\bar 3} {6^{\prime}}}{3} + \sqrt{\frac{3}{7}}\cov{\emb {\bar 3} 6}{\emb 3 {6^{\prime}}}{3}-\\
 &  & -2\sqrt{\frac{2}{21}}\cov{\emb 3 6}{\emb 3 {6^{\prime}}}{\bar{3}_{s}}+\sqrt{\frac{3}{7}}\cov{\emb {\bar 3} 6}{\emb {\bar 3} {6^{\prime}}}{\bar 3} -\sqrt{\frac{2}{21}}\cov{\emb 3  6} {\emb {\bar 3} {6^{\prime}}}{\bar 3} -\sqrt{\frac{2}{21}}\cov{\emb {\bar 3} 6} {\emb 3 {6^{\prime}}}{\bar 3}
\end{eqnarray*}

\item $\mathbf{6}\otimes\mathbf{6}^{\prime}\to\mathbf{7}:$
\begin{eqnarray*}
\mathbf{7} & = & -\frac{1}{\sqrt{2}}\cov{\emb 3 6}{\emb {\bar 3} {6^{\prime}}}{1} + \frac{1}{\sqrt{2}}\cov{\emb {\bar 3} 6} {\emb 3 {6^{\prime}}}{1}- \\ 
& & -\frac{1}{\sqrt{3}}\cov{\emb {\bar3} 6}{\emb {\bar 3} {6^{\prime}}}{3_{a}}-\frac{1}{\sqrt{3}}\cov{\emb 3 6} {\emb {\bar 3} {6^{\prime}}}{3}+\frac{1}{\sqrt{3}}\cov{\emb {\bar 3} 6}{\emb 3 {6^{\prime}}}{3}+\\
&  & +\frac{1}{\sqrt{3}}\cov{\emb 3 6}{\emb 3 {6^{\prime}}}{\bar{3}_{a}}-\frac{1}{\sqrt{3}}\cov{\emb 3  6}{\emb {\bar 3} {6^{\prime}}}{\bar 3}+\frac{1}{\sqrt{3}}\cov{\emb {\bar 3} 6}{\emb 3 {6^{\prime}}}{\bar 3}
\end{eqnarray*}

\end{itemize}

\vskip .3cm

\newpage
\section{Building $\m M$ with $\m T_7$ Invariants}

\noindent As mentioned in the main text, aside from the linear combination Eq.~\ref{eq:lc}, there exist other dimension-five $T_{7}$ invariants capable
of producing the special Majorana matrix 
\begin{equation}
\mathcal{M}=\left(\begin{array}{ccc}
r\lambda^{16} & r\lambda^{12} & r\lambda^{8}\\
r\lambda^{12} & \lambda^{8} & -\lambda^{4}\\
r\lambda^{8} & -\lambda^{4} & 1
\end{array}\right).\label{eq:App-inv0-smm}
\end{equation}
 Here, we find all such dimension-five $T_{7}$ invariants built out
of two right-handed neutrino fields and two familon fields. The right-handed
neutrinos are $T_{7}$ triplets $\left(N\sim\rep{3}\right)$ and the
familon fields are either triplets $\left(\varphi,\varphi^{\prime}\sim\rep{3}\right)$
or antitriplets $\left(\bar{\varphi},\bar{\varphi}^{\prime}\sim\rep{\bar{3}}\right)$.
These $T_{7}$ invariants fall into three classes by the representations
of the two familon fields \cite{Majorana_Paper}.
\begin{itemize}
\item The $NN\varphi\varphi^{\prime}$ type invariants contain two linearly
independent invariants: 
\begin{eqnarray*}
\mathcal{I}_{1}^{\left(\varphi,\varphi'\right)} & = & \scov{{N}\,{N}}{3}\scov{\varphi\varphi'}{\bar{3}_{s}}-\scov{N\, N}{3_{s}}\scov{\varphi\varphi'}{\bar{3}_{a}}.\\
 & \longrightarrow & \frac{1}{\sqrt{6}}\left(\begin{array}{ccc}
\varphi_{1}\varphi_{3}' & 0 & 0\\
0 & \varphi_{2}\varphi_{1}' & 0\\
0 & 0 & \varphi_{3}\varphi_{2}'
\end{array}\right),\\
\mathcal{I}_{2}^{\left(\varphi,\varphi'\right)} & = & \scov{{N}\,{N}}{\bar{3}}\scov{\varphi\varphi'}{3}\\
 & \longrightarrow & \frac{1}{\sqrt{6}}\left(\begin{array}{ccc}
0 & \varphi_{2}\varphi_{2}' & \varphi_{1}\varphi_{1}'\\
\varphi_{2}\varphi_{2}' & 0 & \varphi_{3}\varphi_{3}'\\
\varphi_{1}\varphi_{1}' & \varphi_{3}\varphi_{3}' & 0
\end{array}\right).
\end{eqnarray*}

\item The $NN\varphi\bar{\varphi}$ type invariants contain two linearly
independent invariants:
\begin{eqnarray*}
\mathcal{J}_{1}^{\left(\varphi,\bar{\varphi}\right)} & = & \scov{{N}\,{N}}{3}\scov{\varphi\bar{\varphi}}{\bar{3}}\longrightarrow\frac{1}{\sqrt{3}}\left(\begin{array}{ccc}
\varphi_{2}\bar{\varphi}_{3} & 0 & 0\\
0 & \varphi_{3}\bar{\varphi}_{1} & 0\\
0 & 0 & \varphi_{1}\bar{\varphi}_{2}
\end{array}\right),\\
\mathcal{J}_{2}^{\left(\varphi,\bar{\varphi}\right)} & = & \scov{{N}\,{N}}{\bar{3}}\scov{\varphi\bar{\varphi}}{3}\longrightarrow\frac{1}{\sqrt{6}}\left(\begin{array}{ccc}
0 & \varphi_{1}\bar{\varphi}_{3} & \varphi_{3}\bar{\varphi}_{2}\\
\varphi_{1}\bar{\varphi}_{3} & 0 & \varphi_{2}\bar{\varphi}_{1}\\
\varphi_{3}\bar{\varphi}_{2} & \varphi_{2}\bar{\varphi}_{1} & 0
\end{array}\right).
\end{eqnarray*}

\item The $NN\bar{\varphi}\bar{\varphi}^{\prime}$ type invariants contain
two linearly independent invariants:
\begin{eqnarray*}
\mathcal{K}_{1}^{\left(\bar{\varphi},\bar{\varphi}^{\prime}\right)} & = & \scov{{N}\,{N}}{3}\scov{\bar{\varphi}\bar{\varphi}^{\prime}}{\bar{3}}\longrightarrow\frac{1}{\sqrt{3}}\left(\begin{array}{ccc}
\bar{\varphi}_{1}\bar{\varphi}_{1}^{\prime} & 0 & 0\\
0 & \bar{\varphi}_{2}\bar{\varphi}_{2}^{\prime} & 0\\
0 & 0 & \bar{\varphi}_{3}\bar{\varphi}_{3}^{\prime}
\end{array}\right),\\
\mathcal{K}_{2}^{\left(\bar{\varphi},\bar{\varphi}^{\prime}\right)} & = & \scov{{N}\,{\bar{\varphi}}}{3}\scov{N\bar{\varphi}^{\prime}}{\bar{3}}\longrightarrow\frac{1}{2\sqrt{3}}\left(\begin{array}{ccc}
0 & \bar{\varphi}_{1}\bar{\varphi}_{2}^{\prime} & \bar{\varphi}_{3}\bar{\varphi}_{1}^{\prime}\\
\bar{\varphi}_{1}\bar{\varphi}_{2}^{\prime} & 0 & \bar{\varphi}_{2}\bar{\varphi}_{3}^{\prime}\\
\bar{\varphi}_{3}\bar{\varphi}_{1}^{\prime} & \bar{\varphi}_{2}\bar{\varphi}_{3}^{\prime} & 0
\end{array}\right).
\end{eqnarray*}

\end{itemize}
The details of the calculation for each class are as follows.

\subsection*{$NN\varphi\varphi^{\prime}$ Invariants}

The general form of the $NN\varphi\varphi^{\prime}$ invariant is
\[
\mathcal{I}=\mathcal{I}_{2}+a\mathcal{I}_{1}^{\left(\varphi,\varphi^{\prime}\right)}+b\mathcal{I}_{1}^{\left(\varphi^{\prime},\varphi\right)},
\]
 where $a,b$ are two constants. Parametrizing the familon vacuum values as
\begin{equation}
\left\langle \varphi\right\rangle \sim\left\langle \varphi_{3}\right\rangle \left(\begin{array}{c}
\lambda^{m}\\
\lambda^{n}\\
1
\end{array}\right),\quad\left\langle \varphi'\right\rangle \sim\left\langle \varphi'_{3}\right\rangle \left(\begin{array}{c}
\lambda^{m'}\\
\lambda^{n'}\\
1
\end{array}\right),\label{eq:App-Inv1-vacuum_1}
\end{equation}
 and substituing them into $\mathcal{I}$, we obtain
\[
\left\langle \mathcal{I}\right\rangle \sim\left(\begin{array}{ccc}
\lambda^{\mathrm{min}(m,m')} & \lambda^{n+n'} & \lambda^{m+m'}\\
\lambda^{n+n'} & \lambda^{\mathrm{min}(m+n',n+m')} & 1\\
\lambda^{m+m'} & 1 & \lambda^{\mathrm{min}(n',n)}
\end{array}\right).
\]
Comparing the powers of the $\left(11\right)$ and $\left(13\right)$ elements
between Eq.~\ref{eq:App-inv0-smm} and $\left\langle \mathcal{I}\right\rangle $
yields
\[
\mathrm{min}\left(m,m'\right)=12,\quad m+m^{\prime}=4.
\]
 As these two equations conflict, one of $a,b$ must be zero. Without
loss of generality, we can set $b=0$ and obtain 
\begin{equation}
\left\langle \mathcal{I}\right\rangle \sim\left(\begin{array}{ccc}
\lambda^{m} & \lambda^{n+n'} & \lambda^{m+m'}\\
\lambda^{n+n'} & \lambda^{n+m'} & 1\\
\lambda^{m+m'} & 1 & \lambda^{n'}
\end{array}\right).\label{eq:App-Inv1-Ivev}
\end{equation}
 Comparing the powers of $\lambda$ between Eq.~\ref{eq:App-inv0-smm}
and Eq.~\ref{eq:App-Inv1-Ivev} yields
\[
m=n=12,\quad m^{\prime}=-8,\quad n^{\prime}=-4.
\]
With the above solution, we can parametrize the familon vacuum as
\[
\left\langle \varphi\right\rangle =\left\langle \varphi_{3}\right\rangle \left(\begin{array}{c}
\alpha\lambda^{12}\\
\beta\lambda^{12}\\
1
\end{array}\right),\quad\left\langle \varphi'\right\rangle =\left\langle \varphi'_{1}\right\rangle \left(\begin{array}{c}
1\\
\beta'\lambda^{4}\\
\gamma'\lambda^{8}
\end{array}\right),
\]
 and obtain 
\[
\left\langle \mathcal{I}\right\rangle =\left\langle \varphi_{3}\right\rangle \left\langle \varphi'_{1}\right\rangle \lambda^{4}\left(\begin{array}{ccc}
\alpha\beta'\lambda^{16} & a\beta\beta'\lambda^{12} & a\alpha\lambda^{8}\\
a\beta\beta'\lambda^{12} & \beta\lambda^{8} & a\gamma'\lambda^{4}\\
a\alpha\lambda^{8} & a\gamma'\lambda^{4} & \beta'
\end{array}\right).
\]
Comparing the prefactor of $\left\langle \mathcal{I}\right\rangle $
and Eq.~\ref{eq:App-inv0-smm} yields
\begin{equation}
a=-r^{-\frac{1}{3}},\quad\left\langle \varphi\right\rangle =\left\langle \varphi_{3}\right\rangle \left(\begin{array}{c}
r^{\frac{4}{3}}\lambda^{12}\\
-r^{\frac{2}{3}}\lambda^{12}\\
1
\end{array}\right),\quad\left\langle \varphi'\right\rangle =\left\langle \varphi_{1}^{\prime}\right\rangle \left(\begin{array}{c}
1\\
-r^{\frac{2}{3}}\lambda^{4}\\
r^{\frac{1}{3}}\lambda^{8}
\end{array}\right).\label{eq:App-Inv1-Vacum_1}
\end{equation}
 The only invariant capable of producing the special Majorana
matrix is therefore
\begin{equation}
\mathcal{I}=\mathcal{I}_{2}-r^{\frac{1}{3}}\mathcal{I}_{1}^{\left(\varphi,\varphi'\right)},\label{eq:App-Inv1-I}
\end{equation}
 which depends on the value of $r$.

\subsection*{$NN\varphi\bar{\varphi}$ Invariants}

The general form of the $NN\varphi\bar{\varphi}$ invariant is
\[
\mathcal{J}=\mathcal{J}_{1}^{\left(\varphi\bar{\varphi}\right)}+a\mathcal{J}_{2}^{\left(\varphi,\bar{\varphi}\right)}.
\]
Parametrizing the familon vacuum as
\[
\left\langle \varphi\right\rangle =\left\langle \varphi_{1}\right\rangle \left(\begin{array}{c}
1\\
\alpha\lambda^{n}\\
\beta\lambda^{m}
\end{array}\right),\quad\left\langle \bar{\varphi}\right\rangle =\left\langle \bar{\varphi}_{1}\right\rangle \left(\begin{array}{c}
1\\
\bar{\alpha}\lambda^{\bar{n}}\\
\bar{\beta}\lambda^{\bar{m}}
\end{array}\right),
\]
and going through similar calculations of the previous subsection,
we obtain
\begin{equation}
\left\langle \varphi\right\rangle =\left\langle \varphi_{1}\right\rangle \left(\begin{array}{c}
1\\
-r^{\frac{1}{3}}\lambda^{4}\\
-r^{\frac{2}{3}}\lambda^{8}
\end{array}\right),\quad\left\langle \bar{\varphi}\right\rangle =\left\langle \bar{\varphi}_{1}\right\rangle \left(\begin{array}{c}
1\\
-r^{\frac{2}{3}}\lambda^{4}\\
-r^{\frac{4}{3}}\lambda^{12}
\end{array}\right),\label{eq:App-Inv2-Vacuum_1}
\end{equation}
 and
\[
a=-\sqrt{2}r^{\frac{1}{3}}.
\]
Therefore, the only operator capable of generating the special Majorana
matrix is the $r$-dependent linear combination
\begin{equation}
\mathcal{J}=\mathcal{J}_{1}-\sqrt{2}r^{\frac{1}{3}}\mathcal{J}_{2}^{\left(\varphi,\bar{\varphi}\right)}.\label{eq:App-Inv2-invariant}
\end{equation}

\subsection*{$NN\bar{\varphi}\bar{\varphi}^{\prime}$ Invariants}

The general form of the $NN\bar{\varphi}\bar{\varphi}^{\prime}$ invariant is 
\begin{equation}
\mathcal{K}=\mathcal{K}_{1}^{\left(\bar{\varphi},\bar{\varphi}^{\prime}\right)}+a\mathcal{K}_{2}^{\left(\bar{\varphi},\bar{\varphi}^{\prime}\right)}+b\mathcal{K}_{2}^{\left(\bar{\varphi}^{\prime},\bar{\varphi}\right)}.\label{eq:App-Inv3-K1}
\end{equation}
We parametrize the familon vacuum as
\[
\left\langle \bar{\varphi}\right\rangle =\left\langle \bar{\varphi}_{3}\right\rangle \left(\begin{array}{c}
\bar{\alpha}\lambda^{m}\\
\bar{\beta}\lambda^{n}\\
1
\end{array}\right),\quad\left\langle \bar{\varphi}^{\prime}\right\rangle =\left\langle \bar{\varphi}_{3}^{\prime}\right\rangle \left(\begin{array}{c}
\bar{\alpha}^{\prime}\lambda^{m^{\prime}}\\
\bar{\beta}^{\prime}\lambda^{n^{\prime}}\\
1
\end{array}\right).
\]
To match the powers of $\lambda$ in the special Majorana matrix, it
must have $m=m^{\prime}=8,n=n^{\prime}=4$. Feeding the familon vacuum
into $\mathcal{K}$ yields 
\begin{equation}
\left\langle \mathcal{K}\right\rangle \sim\left(\begin{array}{ccc}
\bar{\alpha}\bar{\alpha}'\lambda^{16} & \frac{1}{2}\left(b\bar{\alpha}'\bar{\beta}+a\bar{\alpha}\bar{\beta}'\right)\lambda^{12} & \frac{1}{2}\left(b\bar{\alpha}+a\bar{\alpha}'\right)\lambda^{8}\\
\frac{1}{2}\left(b\bar{\alpha}'\bar{\beta}+a\bar{\alpha}\bar{\beta}'\right)\lambda^{12} & \bar{\beta}\bar{\beta}'\lambda^{8} & \frac{1}{2}\left(a\bar{\beta}+b\bar{\beta}'\right)\lambda^{4}\\
\frac{1}{2}\left(b\bar{\alpha}+a\bar{\alpha}'\right)\lambda^{8} & \frac{1}{2}\left(a\bar{\beta}+b\bar{\beta}'\right)\lambda^{4} & 1
\end{array}\right)\label{eq:App-Inv3-K-vev}
\end{equation}
Comparing the prefactors between $\left\langle \mathcal{K}\right\rangle $
and the special Majorana matrix, we obtain the following equations
\begin{equation}
\begin{cases}
\bar{\alpha}\bar{\alpha}' & =r\\
a\bar{\alpha}\bar{\beta}'+b\bar{\alpha}'\bar{\beta} & =2r\\
b\bar{\alpha}+a\bar{\alpha}' & =2r\\
\bar{\beta}\bar{\beta}' & =1\\
a\bar{\beta}+b\bar{\beta}' & =-1
\end{cases}\label{eq:App-Inv3-eqs1}
\end{equation}
Considering them as equations for $\bar{\alpha},\bar{\alpha}^{\prime},\bar{\beta},\bar{\beta}^{\prime}$,
we have five equations with only four variables. In general, there
is no solution unless some of the equations are dependent. The only
way to make them solvable is to make the second and third equations
of Eqs. \ref{eq:App-Inv3-eqs1} equivalent,
\begin{eqnarray*}
a\bar{\beta}^{\prime}=b & \Rightarrow & \bar{\beta}^{\prime}=\frac{b}{a},\\
b\bar{\beta}=a & \Rightarrow & \bar{\beta}=\frac{a}{b}.
\end{eqnarray*}
Eqs. \ref{eq:App-Inv3-eqs1} then
become
\begin{eqnarray*}
\bar{\alpha}\bar{\alpha}' & = & r\\
b\bar{\alpha}+a\bar{\alpha}^{\prime} & = & 2r\\
\frac{a^{2}}{b}+\frac{b^{2}}{a} & = & -2
\end{eqnarray*}
 These equations can be solved with a free parameter $k$ as 
\begin{equation}
a=-\frac{2}{k^{2}+\frac{1}{k}},\quad b=-\frac{2k}{k^{2}+\frac{1}{k}},
\end{equation}
\begin{equation}
\bar{\alpha}\bar{\alpha}^{\prime}=r,\quad\bar{\alpha}+k\bar{\alpha}^{\prime}=-r\left(k^{2}+\frac{1}{k}\right).
\end{equation}
This solution gives the familon vacuum
\begin{equation}
\left\langle \bar{\varphi}\right\rangle =\left\langle \bar{\varphi}_{3}\right\rangle \left(\begin{array}{c}
\bar{\alpha}\lambda^{8}\\
\frac{1}{k}\lambda^{4}\\
1
\end{array}\right),\quad\left\langle \bar{\varphi}^{\prime}\right\rangle =\left\langle \bar{\varphi}_{3}^{\prime}\right\rangle \left(\begin{array}{c}
\bar{\alpha}^{\prime}\lambda^{8}\\
k\lambda^{4}\\
1
\end{array}\right).
\end{equation}
For the special case of $k=1$, we get $a=b=-1$ and
\begin{equation}
\mathcal{K}=\mathcal{K}_{1}^{\left(\bar{\varphi},\bar{\varphi}^{\prime}\right)}-\mathcal{K}_{2}^{\left(\bar{\varphi},\bar{\varphi}^{\prime}\right)}-\mathcal{K}_{2}^{\left(\bar{\varphi}^{\prime},\bar{\varphi}\right)},\label{eq:App-Inv3-K2}
\end{equation}
 which is just the matrix generated by the linear combination in the
main text
\begin{equation}
\scov{{\bar{N}}\,{\bar{N}}}{3}\scov{\bar{\varphi}\bar{\varphi}'}{\bar{3}}-\scov{\bar{N}\,\bar{N}}{\bar{3}_{s}}\scov{\bar{\varphi}\bar{\varphi}'}{3_{s}},\label{eq:App-Inv3-Linear-Comb}
\end{equation}
 and the corresponding vacuum values are 
\[
\left\langle \bar{\varphi}\right\rangle =\left\langle \bar{\varphi}_{3}\right\rangle \left(\begin{array}{c}
\bar{\alpha}\lambda^{8}\\
\lambda^{4}\\
1
\end{array}\right),\quad\left\langle \bar{\varphi}^{\prime}\right\rangle =\left\langle \bar{\varphi}_{3}^{\prime}\right\rangle \left(\begin{array}{c}
\bar{\alpha}^{\prime}\lambda^{8}\\
\lambda^{4}\\
1
\end{array}\right),
\]
with $\bar{\alpha}\bar{\alpha}^{\prime}=-\frac{1}{2}\left(\bar{\alpha}+\bar{\alpha}^{\prime}\right)=r$.
The linear combination of Eq.~\ref{eq:App-Inv3-Linear-Comb} is $r$-independent
and contains only two invariants. We find no pair $(a,b)$
capable of converting $\mathcal{K}$ to a single invariant. In this sense,
our choice of the linear combination ~\ref{eq:App-Inv3-Linear-Comb}
is the simplest one. 

\newpage
\section{$\bs\mu$-matrix eigenvectors and eigenvalues}

\noindent The $\bf 6_1$ coupling generates the $\mu$ matrix
\bea
{\large {\bs\mu}}~=~M^{\prime}_0\,\begin{pmatrix}
-r\lambda^{16} &r\lambda^{12} & r\lambda^{8}&0&\sqrt{2}\lambda^4&0 \cr
  r\lambda^{12} & -\lambda^{8} &-\lambda^{4}&0&0&-\sqrt{2}r\lambda^8 \cr
r\lambda^8 &  -\lambda^4& -1&-\sqrt{2}r\lambda^{12}&0&0\cr
0&0&-\sqrt{2}r\lambda^{12}&0&\lambda^8&r\lambda^{16}\cr
\sqrt{2}\lambda^4&0&0&\lambda^8&0&1\cr
0&-\sqrt{2}r\lambda^8&0&r\lambda^{16}&1&0\end{pmatrix},
\nonumber
\eea
whose $\mu$-values and eigenvectors are:
\bea \mu^{}_{E^{(1)}}&=&-M^{\prime}_0(1+\lambda^{8}+r^2\lambda^{16}+4r^2\lambda^{24}+\cdots)\nonumber\\
\noalign{\smallskip}
E^{(1)}_{}&=&-\frac{3r+1}{4}\lambda^8\,H^{}_{u1}+\lambda^4H^{}_{u2}+H^{}_{u3}+\frac{9r-1}{4\sqrt{2}}\lambda^{12}\,\overline{H}^{}_{u1}+\frac{1-r}{4\sqrt{2}}\lambda^4\,\overline{H}^{}_{u2}-\frac{1-r}{4\sqrt{2}}\lambda^4\,\overline{H}^{}_{u3}
\nonumber\\
\noalign{\bigskip} 
\mu^{}_{E^{(2)}}&=&-M^{\prime}_0(1+\lambda^{8}+r^2\lambda^{16}-4r^2\lambda^{24}+\cdots)\nonumber\\
\noalign{\smallskip}
{E^{(2)}_{}}&=&\lambda^4\,H^{}_{u1}+\frac{3r+1}{4}\lambda^8\,H^{}_{u2}+\frac{1-r}{4}\lambda^4\,H^{}_{u3}+\frac{1}{\sqrt{2}}\lambda^8\,\overline{H}^{}_{u1}-\frac{1}{\sqrt{2}}\,\overline{H}^{}_{u2}+\frac{1}{\sqrt{2}}\,\overline{H}^{}_{u3}
\nonumber\\
\noalign{\bigskip} 
\mu^{}_{E^{(3)}}&=&M^{\prime}_0(1+\lambda^{8}+r^2\lambda^{16}-4r^2\lambda^{24}+\cdots)\nonumber\\
\noalign{\smallskip}
E^{(3)}_{}&=&\lambda^4\,H^{}_{u1}-r\lambda^8\,H^{}_{u2}+r\lambda^{12}\,H^{}_{u3}+\frac{1}{\sqrt{2}}\lambda^8\,\overline{H}^{}_{u1}+\frac{1}{\sqrt{2}}\,\overline{H}^{}_{u2}+\frac{1}{\sqrt{2}}\,\overline{H}^{}_{u3}
\nonumber\\
\noalign{\bigskip}
 \mu^{}_{D^{(1)}}&=& M^{\prime}_0(2r\lambda^{12}+\frac{r(r-1)}{2}\lambda^{16}+ \cdots)\nonumber\\
D^{(1)}_{}&=&\frac{1}{\sqrt{2}}\,H^{}_{u1}+\frac{1}{\sqrt{2}}\,H^{}_{u2}-\frac{1}{\sqrt{2}}\lambda^{4}\,H^{}_{u3}+\lambda^4\,\overline{H}^{}_{u1}+r\lambda^8\,\overline{H}^{}_{u2}-\lambda^4\,\overline{H}^{}_{u3}
\nonumber\\
\noalign{\bigskip}
\mu^{}_{D^{(2)}}&=& M^{\prime}_0(-2r\lambda^{12}+\frac{r(r-1)}{2}\lambda^{16}+ \cdots)\nonumber\\
\noalign{\smallskip}
D^{(2)}_{}&=&-\frac{1}{\sqrt{2}}\,H^{}_{u1}+\frac{1}{\sqrt{2}}\,H^{}_{u2}-\frac{1}{\sqrt{2}}\lambda^{4}\,H^{}_{u3}-\lambda^4\,\overline{H}^{}_{u1}+r\lambda^8\,\overline{H}^{}_{u2}+\lambda^4\,\overline{H}^{}_{u3}
\nonumber\\
\noalign{\bigskip}
\mu^{}_{h}&= &8r^2\lambda^{24}M^{\prime}_0(1+ \cdots)\nonumber\\
\noalign{\smallskip}\noalign{\smallskip}
h &=&\sqrt{2}\lambda^4\,H^{}_{u1}-r\sqrt{2}\lambda^8\,H^{}_{u2}+3r\sqrt{2}\lambda^{12}\,H^{}_{u3}-\overline{H}^{}_{u1}+r(1-2r)\lambda^{16}\,\overline{H}^{}_{u2}-\lambda^8\,\overline{H}^{}_{u3}
\nonumber
\eea

\noindent The $\bf 6_2$ coupling generates the $\mu$ matrix 
\bea 
{\large {\bs\mu}} ~=~ M^{\prime\prime}_0\left(\begin{array}{cccccc}
r\lambda^{16} & 2r\lambda^{12} & 2r\lambda^{8} & 0 & -\sqrt{2}\lambda^{4} & -\frac{3\lambda^{8}}{\sqrt{2}}\\
2r\lambda^{12} & \lambda^{8} & -2\lambda^{4} & -\frac{3}{\sqrt{2}} & 0 & \sqrt{2}r\lambda^{8}\\
2r\lambda^{8} & -2\lambda^{4} & 1 & \sqrt{2}r\lambda^{12} & -\frac{3r\lambda^{16}}{\sqrt{2}} & 0\\
0 & -\frac{3}{\sqrt{2}} & \sqrt{2}r\lambda^{12} & -3r\lambda^{8} & \frac{\lambda^{8}}{2} & \frac{r\lambda^{16}}{2}\\
-\sqrt{2}\lambda^{4} & 0 & -\frac{3r\lambda^{16}}{\sqrt{2}} & \frac{\lambda^{8}}{2} & -3r\lambda^{12} & \frac{1}{2}\\
-\frac{3\lambda^{8}}{\sqrt{2}} & \sqrt{2}r\lambda^{8} & 0 & \frac{r\lambda^{16}}{2} & \frac{1}{2} & 3\lambda^{4}
\end{array}\right),
\nonumber
\eea
where $M^{\prime\prime}_0=\frac{f}{\sqrt{7}}\times 10^{14}$ \rm{GeV}. The $\mu$-values and eigenvectors are:
\bea \mu_{1} &=& M^{\prime\prime}_0(1-\frac{8}{7}\lambda^{8}+ \cdots)\nonumber\\
\noalign{\smallskip}
E^{\left(1\right)} &=& 2r\lambda^{8}H_{u1}+\frac{4}{7}\lambda^{4}H_{u2}+H_{u3}-\frac{6\sqrt{2}}{7}\lambda^{4}\overline{H}_{u1}- \nonumber\\
& & - \frac{4}{7}\sqrt{2}(4r+1)\lambda^{12}\overline{H}_{u2}-\frac{2}{7}\sqrt{2}(2r+1)\lambda^{12}\overline{H}_{u3} \nonumber\\
\noalign{\bigskip} 
\mu_{2} &=& M^{\prime\prime}_0\left(-\frac{1}{2}+\frac{3}{2}\lambda^{4}-\frac{17}{4}\lambda^{8}-\frac{3}{2}(r+4)\lambda^{12}+\cdots\right)\nonumber\\
\noalign{\smallskip}
E^{\left(2\right)} &=& 2\lambda^{4}H_{u1}-\frac{1}{17}(2r-3)\lambda^{8}H_{u2}-\frac{4}{17}(12r-1)\lambda^{12}H_{u3} -\nonumber\\
& & -\frac{12r-1}{17\sqrt{2}}\lambda^{8}\overline{H}_{u1}+\frac{1}{\sqrt{2}}\overline{H}_{u2}-\frac{1}{\sqrt{2}}\overline{H}_{u3} \nonumber\\
\noalign{\bigskip} 
\mu_{3} &=& M^{\prime\prime}_0\left(\frac{1}{2}+\frac{3}{2}\lambda^{4}+\frac{17}{4}\lambda^{8}-\frac{3}{2}(r+4)\lambda^{12}+\cdots\right) \nonumber\\
\noalign{\smallskip}
E^{\left(3\right)} &=& -2\lambda^{4}H_{u1}-\frac{1}{17}(2r-3)\lambda^{8}H_{u2}+\frac{4}{17}(32r+3)\lambda^{12}H_{u3} + \nonumber\\
& & + \frac{12r-1}{17\sqrt{2}}\lambda^{8}\overline{H}_{u1}+\frac{1}{\sqrt{2}}\overline{H}_{u2}+\frac{1}{\sqrt{2}}\overline{H}_{u3} \nonumber\\
\noalign{\bigskip} 
\mu_{4} &=& M^{\prime\prime}_0\left(-\frac{3}{\sqrt{2}}-(\frac{3r}{2}-\frac{15}{14}+\frac{6\sqrt{2}}{7})\lambda^{8}+\cdots\right) \nonumber\\
\noalign{\smallskip}
E^{\left(4\right)} &=& -\left(\frac{74r}{357}+\frac{4\sqrt{2}r}{7}+\frac{2}{17}\right)\lambda^{12}H_{u1}+\frac{1}{\sqrt{2}}H_{u2}+\left(\frac{6}{7}-\frac{2\sqrt{2}}{7}\right)\lambda^{4}H_{u3} + \nonumber\\
& & + \frac{1}{\sqrt{2}}\overline{H}_{u1}+\left(\frac{2r}{17}-\frac{3}{17}\right)\lambda^{8}\overline{H}_{u2}-\left(\frac{6\sqrt{2}r}{17}-\frac{1}{17\sqrt{2}}\right)\lambda^{8}\overline{H}_{u3}
\nonumber
\eea

\bea \mu_{5} &=& M^{\prime\prime}_0\left(\frac{3}{\sqrt{2}}-(\frac{3r}{2}-\frac{15}{14}-\frac{6\sqrt{2}}{7})\lambda^{8}+\cdots\right) \nonumber\\
\noalign{\smallskip}
E^{\left(5\right)} &=& \left(\frac{74r}{357}-\frac{4\sqrt{2}r}{7}+\frac{2}{17}\right)\lambda^{12}H_{u1}+\frac{1}{\sqrt{2}}H_{u2}-\left(\frac{6}{7}+\frac{2\sqrt{2}}{7}\right)\lambda^{4}H_{u3} - \nonumber\\
& & - \frac{1}{\sqrt{2}}\overline{H}_{u1}+\left(\frac{2r}{17}-\frac{3}{17}\right)\lambda^{8}\overline{H}_{u2}+\left(\frac{6\sqrt{2}r}{17}-\frac{1}{17\sqrt{2}}\right)\lambda^{8}\overline{H}_{u3} \nonumber\\
\noalign{\bigskip}
\mu_{6} &=& 12\lambda^{12}M^{\prime\prime}_0(1+\cdots) \nonumber\\
\noalign{\smallskip}
E^{\left(6\right)} &=& H_{u1}-3\lambda^{16}H_{u2}-2r\lambda^{8}H_{u3}+\frac{10}{3}\sqrt{2}r\lambda^{12}\overline{H}_{u1}-9\sqrt{2}\lambda^{8}\overline{H}_{u2}+2\sqrt{2}\lambda^{4}\overline{H}_{u3}
\nonumber
\eea

{}

\end{document}